\documentclass[aps,prb,twocolumn,superscriptaddress,showpacs,floatfix,amsmath,amssymb]{revtex4}
\usepackage{graphicx}
\usepackage{color}
\usepackage{xcolor,soul}
\usepackage[normalem]{ulem}

\DeclareMathAlphabet\mathbfcal{OMS}{cmsy}{b}{n}

\newcommand{\br}[0]{ {\bf r} }

\newcommand{\vm}[0]{ {\vec m} }
\newcommand{\bj}[0]{ {\bf j} }
\newcommand{\bJ}[0]{ {\bf J} }

\newcommand{\bA}[0]{ {\bf A}_{\rm G} }

\usepackage{color}
\usepackage{booktabs}
\usepackage{longtable}
\usepackage{bm}% bold math

\def\bea{\begin{eqnarray}}
\def\eea{\end{eqnarray}}
\def\ben{\begin{equation}}
\def\een{\end{equation}}
\def\benu{\begin{enumerate}}
\def\enu{\end{enumerate}}

\def\br{{\bf r}}

\definecolor{Mygrey}{gray}{0.80}
\definecolor{lteal}{rgb}{0.10,0.60,0.70}
\definecolor{dkred}{rgb}{0.80,0.10,0.00}
\begin{document}

\title{Electron localization function for non-collinear spins}

\author{Jacques K. Desmarais}
\email{jacqueskontak.desmarais@unito.it}
\affiliation{Dipartimento di Chimica, Universit\`{a} di Torino, via Giuria 5, 10125 Torino, Italy}

\author{Giovanni Vignale}
\affiliation{Institute for Functional Intelligent Materials, National University of Singapore, 4 Science Drive 2, Singapore 117544}

\author{Kamel Bencheikh}
\affiliation{Setif 1 University-Ferhat Abbas.
Faculty of Sciences. Department of Physics and Laboratory of Quantum Physics and  Dynamical Systems,
Setif, Algeria}

\author{Alessandro Erba}
\affiliation{Dipartimento di Chimica, Universit\`{a} di Torino, via Giuria 5, 10125 Torino, Italy}

\author{Stefano Pittalis}
\email{stefano.pittalis@nano.cnr.it}
\affiliation{Istituto Nanoscienze, Consiglio Nazionale delle Ricerche, Via Campi 213A, I-41125 Modena, Italy}

\date{\today}

\begin{abstract}
Understanding of bonding is key to modelling  materials and predicting properties thereof. A widely adopted indicator of bonds and atomic shells is the electron localization function (ELF). The building blocks of the ELF are also used in the construction of modern density functional approximations.
Here we demonstrate that the ELF  breaks down when applied beyond  regular non-relativistic quantum states. We show that for tackling general non-collinear open-shell solutions, it is essential to 
address both the U(1) gauge invariance --- i.e., invariance under a multiplication by a position dependent phase factor --- and SU(2) gauge invariance --- i.e. invariance under local spin rotations --- {\em conjointly}.  Remarkably, we find that the extended ELF also improves the description of paradigmatic collinear states.
\end{abstract}

\maketitle

{\em Introduction.}~ 
An ongoing effort  in physics and chemistry is
to turn the concept of bonds among atoms into a quantitative modeling tool to handle understanding and prediction in  most general and novel regimes.\cite{Ball2011}  
This quest started before the discovery of quantum mechanics with the Lewis picture of bonding.\cite{Lewis} This was  then
incarnated in the modern approach by the valence bond (VB) theory of Heitler and London.\cite{heitlerlondon1927} Which, in turn,  inspired the 
concept of ``resonance''  by Pauling\cite{pauling1946resonance} and, eventually, models of condensed  matter in terms of
resonant valence bonds by P.W. Anderson\cite{Anderson1987}
and valence shell electron pair repulsion  theory.\cite{gillespie1957inorganic} 

The idea of sharing electrons is also at the base of   molecular orbital theory, which introduces single-particle orbitals delocalized over the molecule. 
The Kohn-Sham (KS) formulation of density functional theory (DFT) is a molecular orbital theory and, currently, is also the most applied computational methodology in chemistry and materials science. \cite{Jones2015} 
At the center of the success of KS-DFT is the exchange correlation (xc) energy functional --- a functional of the particle density.
A pioneering observation by Bader revealed that the Laplacian of the particle density 
can capture the structure of the atomic shells and bonds~\cite{gatti1988effect,Bader94} --- although not always successfully --- and kicked 
off the challenge of describing bonds within a DFT-like modeling.
Two cornerstones of which are  the ``quantum theory of atoms in molecules and crystals'' and the ``electron localization function''.\cite{gatti2005chemical} In this letter we shall focus on the latter.\cite{Savin94}

Following suggestions by Bader,~\cite{BaderXhole1,BaderXhole2} Becke and Edgecombe  
exploited the anti-symmetry of the many-body state to probe the localization of electrons via the information available at the level of a single 
Slater determinant for (globally) {\em collinear} states: if an electron is localized at a given point in space,  another electron  cannot be brought at the same position with the  same spin.
This behavior can be captured via the same-spin pair density\cite{BE90}
$
P_{\sigma \sigma}(\br_1,\br_2) = N(N-1) \int d3 d4... dN~ | \Psi(\br_1 \sigma,\br_2 \sigma,3,...,N)|^2
$
where the many-electron wave function $\Psi(1,2,3,...,N)$, for simplicity, is taken as an $N$-electron Slater determinant; $1,2, ..., N$ stands for the combined spatial plus spin variables 
$(\br_1,\sigma_1), (\br_2,\sigma_2), ... , (\br_N,\sigma_N)$.
Therefore, the  {\em conditional} probability  $P_{\sigma \sigma}(\br_1, \br_2 )/n_\sigma(\br_1)$ --- where $n_\sigma(\br_1)$ is the spin-electron density ---
refers to the event of finding a second electron at $\br_2$ given the first electron at $\br_1$. Restricting  to collinear states, setting $\br_1 = \br$ and $\br_2= \br_1 + \mathbf{u}$, and taking  the spherical average around $\br_1$,  one then readily finds 
$
P_{\sigma\sigma}(\br, u)/n_\sigma(\br)=\frac{1}{3} D_\sigma(\br) u^2 + \cdot\cdot\cdot\;,
$
where:
$
D_\sigma(\br) = \left[ 2 \tau_\sigma(\br) - \frac{1}{4}  \nabla n_\sigma(\br) \cdot   \nabla n_\sigma(\br)/n_\sigma(\br) \right]
$
and 
$
 \tau_\sigma(\br)  = \frac{1}{2} \Sigma_{i=1,N_\sigma} | \nabla \varphi_{i,\sigma}(\br) |^2
$
is the kinetic energy density of the electrons in the $N_\sigma$ single-particle orbitals
$\varphi_{i,\sigma}(\br)$. 
Next, Becke and Edgecombe defined the {spin-dependent} electron localization function  as follows
\ben\label{ELF}
\mathrm{ELF_\sigma}(\br) \equiv \frac{1}{ 1 +  \left[ D_\sigma(\br) / D^{\rm unif}_\sigma(\br) \right]^2 }\;
\een 
where
$ D^{\rm unif}_\sigma(\br) = \frac{3}{5} \left( 6 \pi^2 \right)^{2/3} n^{5/3}_\sigma(\br) $ 
is $ D_\sigma(\br) $ evaluated for a homogeneous electron gas with the same local density of the real system. 
Eq. \eqref{ELF} is a simple scalar quantity 
 between 0 (zero localization) and 1 (full localization); passing through  1/2 (uniform gas-like localization).

In practice, one usually runs a DFT calculations and evaluates the ratio $D(\br)/D^{\rm unif}(\br)$ using the corresponding KS orbitals.

Since its inception, the ELF has become important beyond the purpose of rationalizing chemical structures and synthesis thereof: it provided insights for 
improving  density functional approximations.\cite{Workhorse2009,SCAN,CHEMSCAN,RG1,TASK,de1998n}
The ELF has also been linked to the ability of density functionals to access local information on the spin-entanglement,\cite{PTRV2015}
to properties of the electronic stress tensor,~\cite{Tao} to features of (time-dependent) density functionals.\cite{BMG05,RG2}

Switching to {\em non-collinear} states, 
the single-particle orbitals must be upgraded to two-component spinors --- none of which may point just ``up'' or ``down''. 
The direction of the  magnetization (defined  more below) may change
from point to point. {\em Locally} the magnetization   provides us with a meaningful {\em collective} direction of spin, but may 
 not  represent the configuration of any of the underlying single-particle spinors. The restriction used in the construction of the ELF is eluded, which
 brings up  the question whether it may be extended to work also for non-collinear states.

\begin{figure}[t!]
\centering
\includegraphics[width=8.6cm]{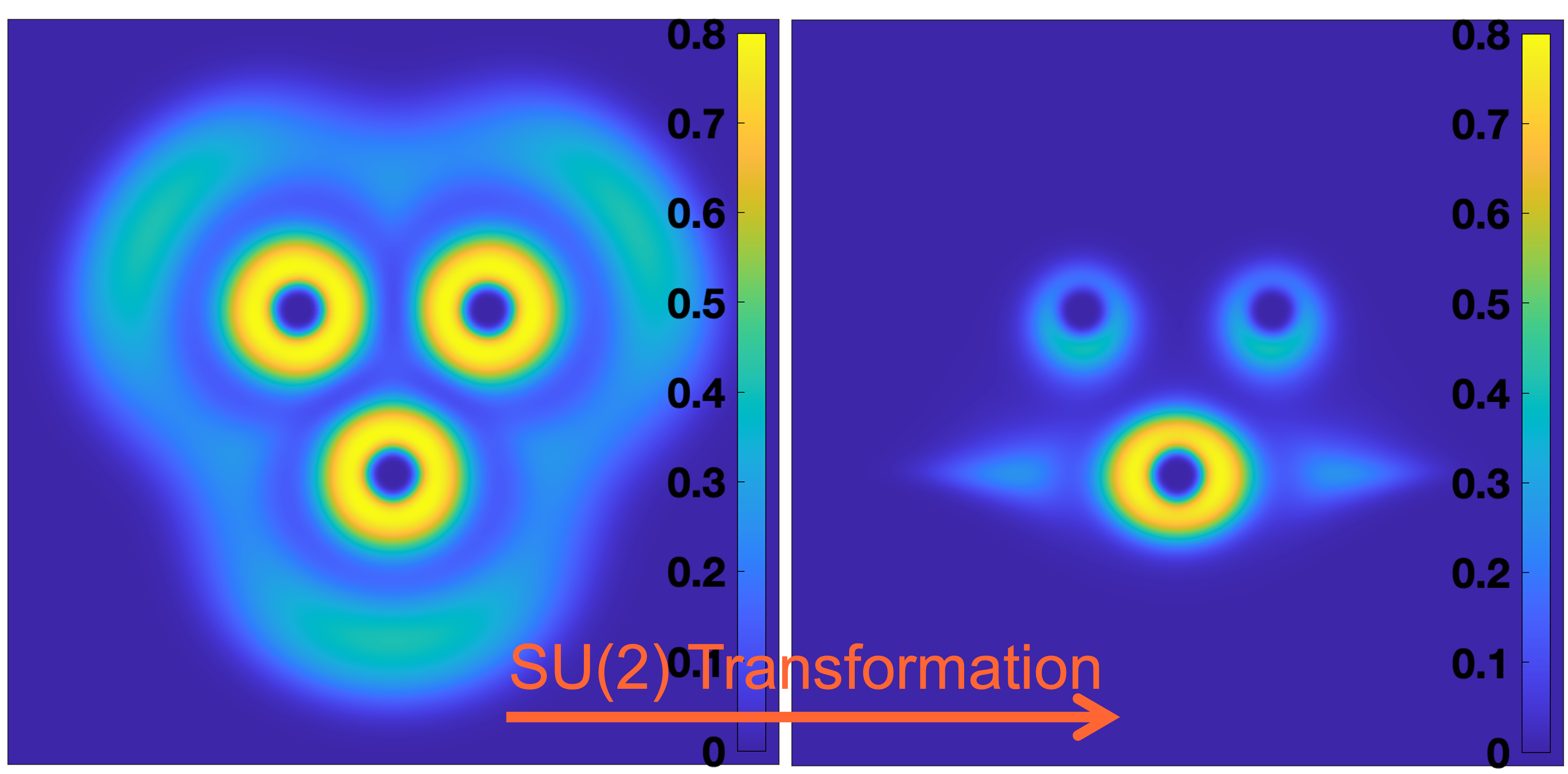}
\caption{Na{\"i}ve ELF of Eq. \eqref{ELFnaive} for the molecular plane of the Cr$_3$ non-collinear magnet obtained before (left) and after (right) an SU(2) gauge transformation (see main text for details). The curious reader can find the plot of the gauge-invariant ELF in Fig. \ref{fig:Cr3_U1SU2}.}\label{fig:Cr3_gauge}
\end{figure}

{\em Solution}. For tackling this problem, let us invoke the expression {\em and} the interpretation of the ELF  by
Savin and collaborators\cite{Savin92} 
\ben\label{ELF2}
\mathrm{ELF}(\br) \equiv \frac{1}{ 1 +  \left[ D(\br) / D^{\rm unif}(\br) \right]^2 }\;
\een 
Here, the spin-resolved quantity $D_\sigma(\br)$ is  replaced by the {\em spin-summed} quantity
\begin{equation}\label{eqn:D}
D(\br) = \tau(\br) - \frac{\nabla n(\br) \cdot   \nabla n(\br)}{8 n(\br)}\;,
\end{equation}
which is thus normalized relative to the spin-unpolarized 
$ D^{\rm unif}(\br) = \frac{3}{10} \left( 3 \pi^2 \right)^{2/3} n^{5/3}(\br) $.
It can be readily shown that\cite{hoffmann1977schrodinger} $D(\br)$ is a non-negative quantity $D(\br) \ge 0$. Thus,  $D(\br)$ can be regarded as the {\em excess} of the fermionic kinetic energy $\tau(\br)$ as compared to the ``bosonic kinetic energy'' $\nabla n(\br) \cdot   \nabla n(\br)/8 n(\br)$: 
small values of the electron localization correspond to large excess fermionic kinetic energy.\cite{Savin92}

Next, initiating the analysis that will bring us to the extension of the ELF, let us consider the gauge aspects involved in the electron localization function.
In modern geometric parlance,\cite{aitchison1989gauge,nakahara2018geometry}  (local) gauge transformations can be understood in terms of the changes in the local frames used in the representation of the quantum states.
This interpretation  makes it apparent that gauge invariance must be a general and stringent condition for any quantity to hold  an intrinsic physical meaning ---  i.e., expressing the independence of a physical property from the particular  choices made for its representation (or computation). Having noted the importance of the generalized gauge principle in guiding the construction of physical models, we now show its disastrous consequences in relation to the ELF.

As a well-known and enlightening example, consider states carrying a (paramagnetic) particle current, $
\bj(\br) =  \sum_{k=1}^{N} \Im\left\{ \varphi^*_k (\br) \left[ \nabla \varphi_k(\br) \right]\right\}$, as for instance, a molecule in an external magnetic field.
The necessary modification of the ELF, can  be derived in several ways.\cite{BMG05,RasanenCastro2008,Furness2016} 
The importance of the dependence of xc functionals  on the particle current has been largely demonstrated
\cite{Dobson93,Becke96,Pittalis2007,Pittalis09,Rasanen09,Oliveira2010,BatesFurche2012,Tricky16,richter2023meta} A unifying approach for adding particle currents to density functionals and the like 
is to consider the effect of a 
U(1) gauge transformations:\cite{TP05} 
$
{\varphi}' ({\bf r})   =    \exp\left[ \frac{i}{c}  \chi(\bf r)  \right] \varphi ({\bf r})\;
$.
The particle density is obviously invariant but the 
kinetic energy density transforms as follows:
\begin{equation}\label{eqn:U1_tau}
\tau_{\rm nc}'(\br) = \tau_{\rm nc}(\br)  + \frac{1}{c} {\bj}(\br) \cdot \bA(\br) + \frac{n(\br)}{2c^2}  \bA(\br) \cdot \bA(\br) \;
\end{equation}
where $\bA (\br) = c \nabla \chi (\br)$ is a pure gauge (Abelian) vector potential. Therefore, a current-dependent and  
gauge invariant  ELF is readily recovered by
replacing $ \tau(\br) \rightarrow  \widetilde{\tau}(\br)= \tau(\br) - \bj(\br)\cdot\bj(\br)/2n(\br)$ in $D(\br)$ in Eq. \eqref{eqn:D}.

Keeping in mind the above observations, let us come back to consider non-collinear (nc)  states, as for instance, the ground state of the planar Cr$_3$ molecular magnet. Ignoring for simplicity the U(1) aspects reviewed just above, it  is tempting to  extend the ELF by simply evaluating 
$\tau(\br)$ on the two-component spinors $\Phi(\br)$: i.e., $\tau (\br) \rightarrow \tau_{\rm nc}(\br)  = \frac{1}{2} \Sigma_{k}  \left[ \nabla \Phi_{k}(\br) \right]^{\dagger} \cdot \nabla \Phi_{k}(\br)$. 
Performing such a replacement in $D(\br)$, yields 
$ D(\br) \rightarrow  D_{\rm nc}(\br) \equiv  \tau(\br) - \frac{1}{8}  \nabla n(\br) \cdot   \nabla n(\br) / n(\br) $.
Which, in turn, would suggest the following na\"ive expression
\begin{equation}\label{ELFnaive} 
\text{ELF}_\text{na\"ive}(\br) = \frac{1}{ 1 +  \left[ D_{\rm nc}(\br) / D^{\rm unif}(\br) \right]^2 }\;.
\end{equation}
Recall now that an SU(2) gauge transformation acts  on the spinors by locally rotating them as follows
${\Phi}' ({\bf r})  = U_{\rm S}(\br) {\Phi} ({\bf r}) = \exp\left[ \frac{i}{c} {\lambda}^a({\bf r}) { \sigma}^a \right]  {\Phi} ({\bf r})\;,$
according to the spin-vector $\vec{\lambda}({\bf r})$. 
It is  straightforward   to verify that under an SU(2) transformation, $\tau_{\rm nc}(\br)$ transforms as follows:
\begin{align}
\label{eqn:SU2_tau}
\tau'_{\rm nc}(\br)  = \tau_{\rm nc}(\br) + \frac{1}{c}   \bJ^a(\br) \cdot \bA^a(\br)
+ \frac{n(\br)}{2c^2}  \bA^a (\br) \cdot \bA^a (\br) \;,
\end{align}
where $
\bJ^a (\br) =  \sum_{k=1}^{N} \Im \left\{ \Phi^\dagger_k (\br){\sigma^a}  \left[ \nabla \Phi_k(\br) \right] \right\}
$ are the (paramagnetic) spin currents and 
$\bA^a(\br) = - \frac{ic}{2}{\rm Tr}\left( {\sigma^a }U_{\rm S}^\dagger(\br)\mathbf{\nabla }U_{\rm S}(\br) \right)$ are the component of a pure gauge (non-Abelian) vector potential.\cite{note_einstein} 

Let us visualize the effect of an SU(2) gauge transformation in the case of the aforementioned  Cr$_3$ molecule. This is 
a planar molecule with C$_{3v}$ symmetry. 
Consider the transformation where
$\frac{1}{c}\lambda^a (\br)= \frac{1}{2} r_{a}^2 \delta_{a,z}$ and, thus, $A^a_\mu(\br)= c  r_{a} \delta_{a,z} \delta_{\mu,z}$ applied to the Hartree-Fock spinorial wavefunction of the system. As a consequence of which, the transformed  na{\"i}ve ELF
is dramatically different from the untransformed image: compare left and right panels of Figure \ref{fig:Cr3_gauge}. 

But a physically meaningful ELF must not change under gauge transformations. In fact, a recent work on time reversal symmetric (TRS) states, has shown that the replacement $\tau_{\rm nc} \to \widetilde{\tau}_{\rm nc}(\br)  = \tau_{\rm nc}(\br) - \bJ^a (\br) \cdot \bJ^a (\br)/2n(\br)$ in $D_{\rm nc}(\br)$, restores invariance under SU(2) transformations, and simultaneously adds a dependence on paramagnetic spin-currents.~\cite{desmarais2024spincurrents} 
Notice, however, that TRS states have vanishing magnetization. Cr$_3$, instead, exhibits a non-collinear magnetic structure. {\em Shouldn't  such a  magnetization play a role in deriving a gauge invariant ELF?}

This above question urges us to examine a so-far neglected aspect:  the effect of 
\textit{conjoined} U(1)$\times$SU(2) transformations! In detail, such a transformation acts on a spinor as follows
$
{\Phi}' ({\bf r}) = \exp \left[ \frac{i}{c}  \chi({\bf r}) {\mathrm I} + \frac{i}{c} {\lambda^a}({\bf r})  {\sigma^a} \right]  {\Phi} ({\bf r})
$.
As a result,
\begin{align}\label{eqn:tau_U1SU2}
\tau_{\rm nc}'(\br) &= \tau_{\rm nc}(\br) + \frac{n(\br)}{2c^2}  \left[ \bA(\br) \cdot \bA(\br) + \bA^a(\br) \cdot \bA^a(\br) \right] \nonumber \\
&+ \frac{1}{c} {\bj}(\br) \cdot \bA(\br) + \frac{1}{c}   \bJ^a(\br) \cdot \bA^a(\br) \nonumber \\
&+ \frac{1}{c^2} {m^a}(\br)  \bA^a(\br) \cdot \bA(\br)  \;.
\end{align}
Eq. \eqref{eqn:tau_U1SU2} is lengthy but  worth a careful consideration. Crucially, we notice that Eq. \eqref{eqn:tau_U1SU2} does not reduce to the  simple sum of terms due to {\em individual} U(1) and SU(2) gauge transformations. In fact, the expression also includes an additional term coming from the {\em coupling} of the U(1) and SU(2) gauge potentials $\bA(\br)$ and $\bA^a(\br)$ via the spin-magnetization $\vec{m}(\br)$. 

Direct cancellation of the transformation-induced terms in Eq. \eqref{eqn:tau_U1SU2}, leads to the gauge invariant quantity
\begin{align}\label{eqn:ttaunc}
\widetilde{\tau}_{\rm nc}(\br) 
&= \left(  
\tau_{\rm nc}(\br) 
- \frac{\bj(\br) \cdot \bj(\br) }{ 2n(\br)} 
+  \frac{\nabla {m}^a(\br) \cdot \nabla {m}^a(\br)  }{8 n(\br)}
\right)
 \nonumber \\
&+ \left( \frac{{m}^a(\br)  {\tau}_{\rm nc}^a(\br)}{n(\br)}   - \frac{ \bJ^a(\br) \cdot \bJ^a(\br)  }{ 2n(\br)  }  \right) 
\;,
\end{align} 
with
$
\vec{{\tau}}_{\rm nc} ({\bf r}) = \frac{1}{2} \sum_{k=1}^N \sum_\mu \Big[  \partial_\mu \Phi^{\dagger}_k({\bf r}) \Big] \vec{\sigma} \partial_\mu \Phi_k({\bf r})
$ being the {\it spin}-kinetic energy density. Eq. \eqref{eqn:ttaunc}
implements the task of enforcing U(1)$\times$SU(2) gauge invariance adding a {\em minimal} set of terms --- only using quantities which involve first-order derivatives. The gauge invariance of Eq. \eqref{eqn:ttaunc} can be more easily verified by exploiting its connection to the curvature of the exchange hole,\cite{Pittalis2017} as shown in Section 1 of the Supplementary Material.\cite{Supp_MAT_ELF} 
Furthermore,
\begin{equation}\label{eqb:tD}
\widetilde{D}_{\rm nc}(\br) \equiv  \widetilde{\tau}_{\rm nc}(\br)  - \frac{1}{8}  \frac{ \nabla n(\br) \cdot   \nabla n(\br) }{ n(\br) }
\end{equation}
is, like $\widetilde{\tau}_{\rm nc}(\br)$, gauge invariant and  is, like $D(\br)$, a non-negative quantity.
Non-negativity is established by writing $\widetilde{D}_{\rm nc}(\br)$  as the trace of a Hermitian positive semi-definite matrix 
in a reference frame in which the spin and particle currents vanish [see Section 2 of the Supplementary Material].
Therefore,  an interpretation emerges naturally:
$\widetilde{D}_{\rm nc}(\br)$ captures the U(1)$\times$SU(2) gauge-invariant {\em excess} of  $\widetilde{\tau}_{\rm nc}(\br)$ as relative to the reference $\frac{1}{8}  \frac{ \nabla n(\br) \cdot   \nabla n(\br) }{ n(\br) }$. Finally, we write the  U(1)$\times$SU(2) gauge-invariant ELF
\ben\label{eqn:ncELF}
\widetilde{\mathrm{ELF}}(\br) \equiv \frac{1}{ 1 +  \left[  \widetilde{D}_{\rm nc}(\br) / D^{\rm unif}(\br) \right]^2 }\;.
\een  

Notice, Eq.~\eqref{eqn:ncELF} reduces to the SU(2) invariant ELF for TRS states,\cite{desmarais2024spincurrents}, it reduces to the U(1) invariant ELF for current-carrying closed shell states,\cite{BMG05,RasanenCastro2008,Furness2016}  and it reduces to the original ELF for zero current closed shell states.\cite{BE90,Savin92} 
Strikingly, Eq.~\eqref{eqn:ncELF} does {\em not} reduce to $D(\br)$ on globally collinear spin-polarized states --- a surprising fact that we shall  carefully consider more below.

{\em Applications and further analysis}. Eq. \eqref{eqn:ncELF} and, thus, Eq. \eqref{eqb:tD} have been implemented in a developer's version of the \textsc{Crystal23} program.\cite{erba2022crystal23}  Computational details are provided in the Supplementary Material. Further analyses are carried out below.

\begin{figure}[t!]
\centering
\includegraphics[width=5.7cm]{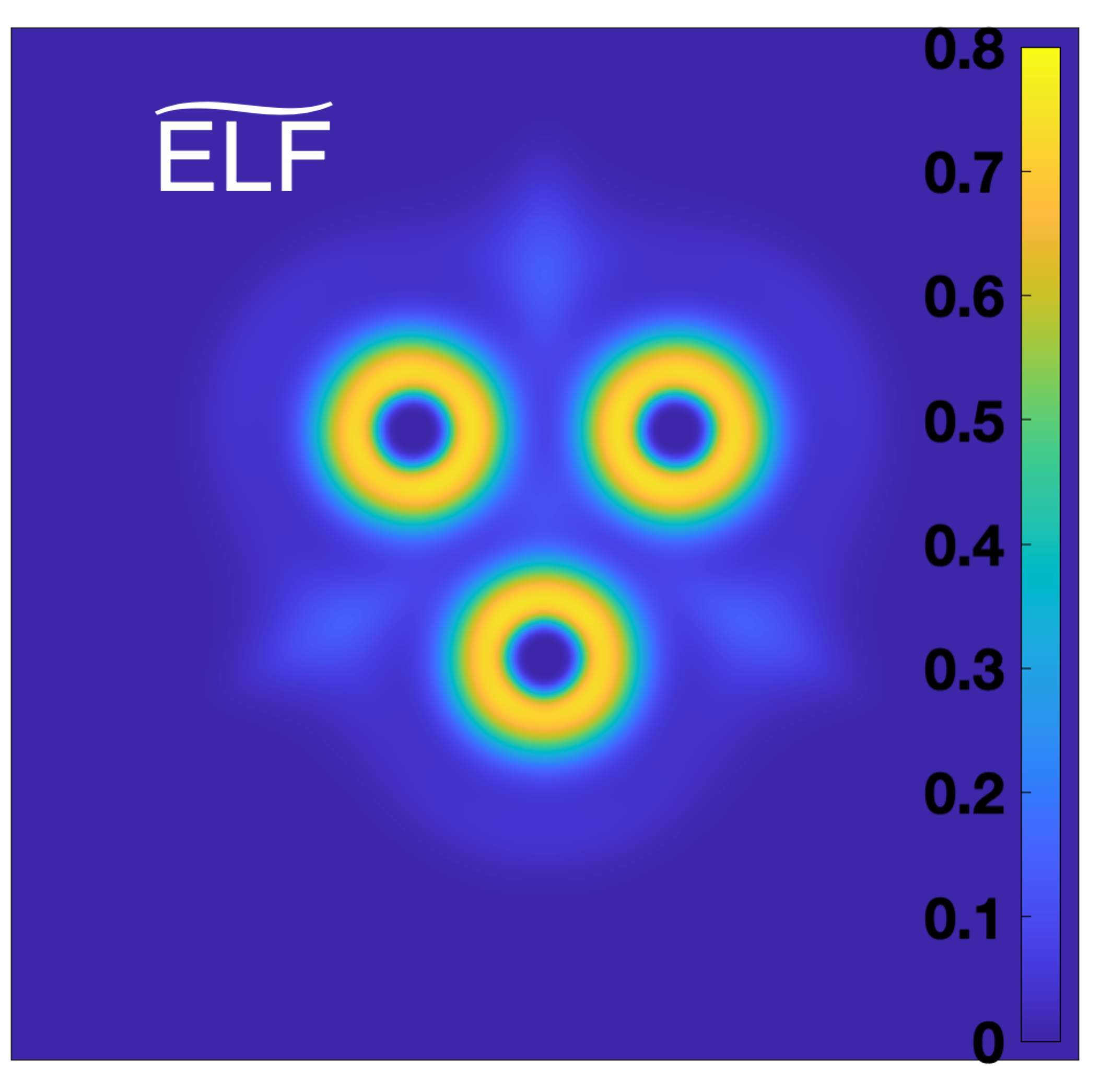}
\caption{U(1)$\times$SU(2) gauge invariant $\widetilde{\mathrm{ELF}}$ [see Eq. \eqref{eqn:ncELF}] in the Cr$_3$ plane.}
\label{fig:Cr3_U1SU2}
\end{figure}

Let us first return to the example of Cr$_3$. The spin-magnetic moments are oriented within the plane of the molecule at angles of 120$^\circ$ in the triangular geometry, which breaks mirror symmetry and leads to a chiral electronic state. Here, the spin-current reaches a magnitude much larger than in any of the other cases discussed below. Particle currents, on the other hand, have much smaller contributions.\cite{cr3_note}

In Fig. \ref{fig:Cr3_U1SU2}, we provide the plot of the U(1)$\times$SU(2) gauge-invariant $\widetilde{\mathrm{ELF}}$ for  Cr$_3$. The non-collinear spin structure increases the conditional pair-probability density at a given point in space, leading to a lower localization in the $\widetilde{\mathrm{ELF}}$, compared to the na\"ive (gauge-dependent) ELF of Eq. \eqref{ELFnaive}. Differences are largest in the anti-bonding region, where the na{\"i}ve ELF significantly overestimates electron localization. In the central bonding region, the largest contribution to the difference $\widetilde{\mathrm{ELF}}$-ELF is provided by the $( \nabla m^a(\br)  )   ( \nabla m^a(\br)  )/8 n$ term. The difference reaches a value as large as $-0.4$, or $40\%$ of the total. 

\begin{figure}[t!]
\centering
\includegraphics[width=8.6cm]{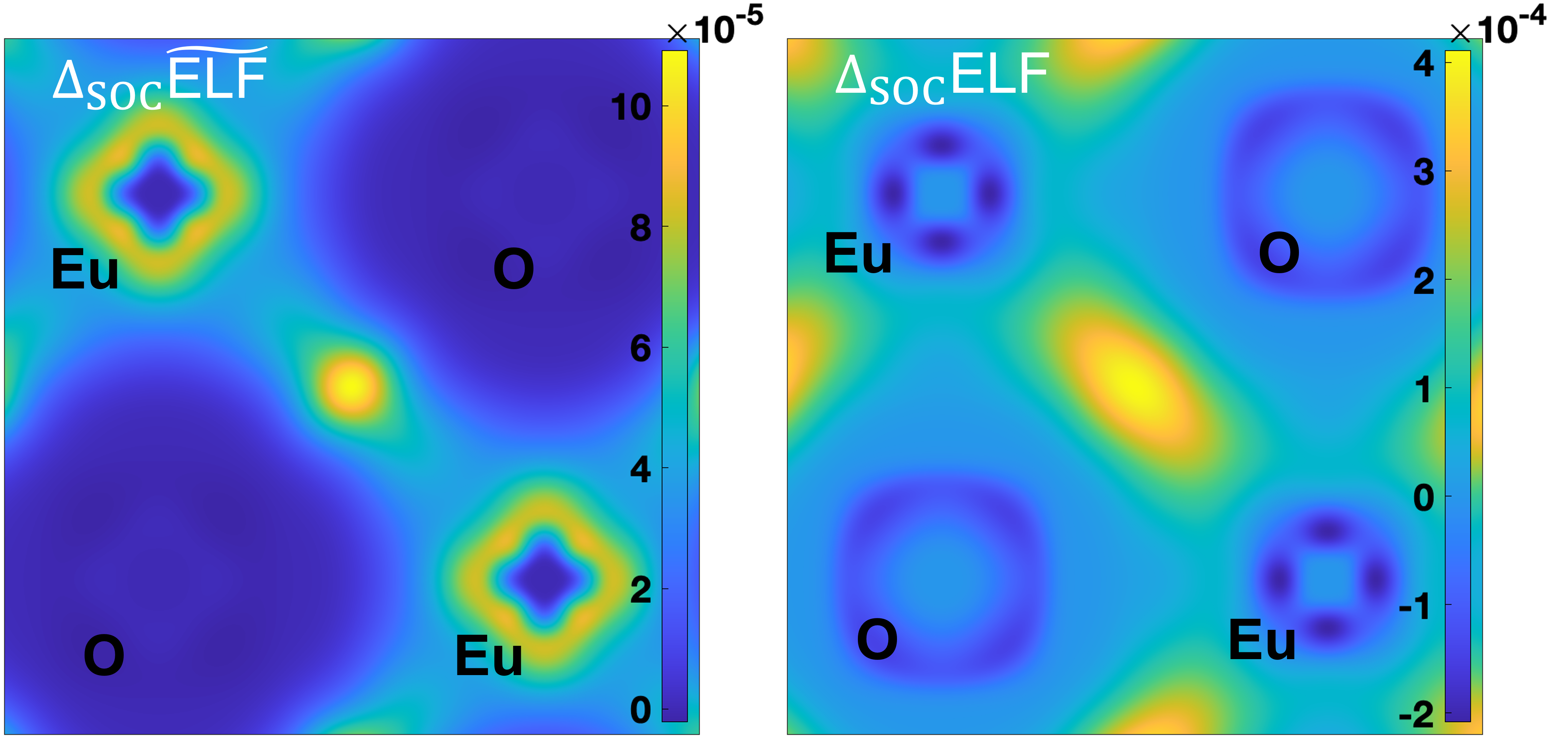}
\caption{(left) Effect of SOC on the U(1)$\times$SU(2) gauge invariant $\widetilde{\mathrm{ELF}}$ [see Eq. \eqref{eqn:ncELF}] in the EuO ferromagnet and (right) the effect of SOC as described by the na{\"i}ve ELF of Eq. \eqref{ELFnaive}. Eu centers are located on the top-left and bottom-right corners, while O centers are located on the top-right and bottom-left corners.}
\label{fig:EuO}
\end{figure}

Let us now look to infinite periodic systems. We consider the prototypical strongly-correlated and mixed-valence material EuO in the insulating phase. The presence of seven ideally spin-filtered $f$ electrons on each Eu atom has attracted significant interest for spintronic applications.\cite{cortes2019tunable,zhang2018high,pradip2016lattice,klinkhammer2014spin} In EuO,  hybridization of the highly localized $d$ and $f$ bands with $s$ and $p$-type ones leads to changes in the bonding picture that may be effectively modulated by external stimuli, such as pressure.\cite{desmarais2021mechanisms,souza2012reentrant} But the effect of SOC on the bonding is yet to be elucidated. We performed global-hybrid density functional theory calculations on the ambient pressure insulating cubic phase of EuO, with SOC included, and all other computational details reported in Ref. \onlinecite{desmarais2021mechanisms}. Inclusion of SOC in the calculation reduces the Kohn-Sham gap from 0.5 to 0.2 eV.

The contribution of the valence band to the $\widetilde{\mathrm{ELF}}$ in EuO is reported in Fig.~S1, along with a plot of the sizeable noncollinear spin-current. The effect of SOC on the bonding is shown through our U(1)$\times$SU(2) gauge-invariant $\widetilde{\mathrm{ELF}}$ in the left panel of Fig. \ref{fig:EuO}, while the right panel reports the effect of SOC on the na{\"i}ve ELF. The plot shows that the contribution to  $\widetilde{\mathrm{ELF}}$ is mostly localized on the heavy Eu atoms, where SOC is large, and at the center of the Eu-Eu direction. In contrast, the gauge-dependent ELF plot is unexpectedly delocalized and does not capture that SOC effects should be larger on the Eu atoms than on the O atoms --- Even worse, different pictures, for physically equivalent gauges, could be generated (as we have shown above for Cr$_3$). Instead, the $\widetilde{\mathrm{ELF}}$ is determined unambiguously. 

\begin{figure}[t!]
\centering
\includegraphics[width=7cm]{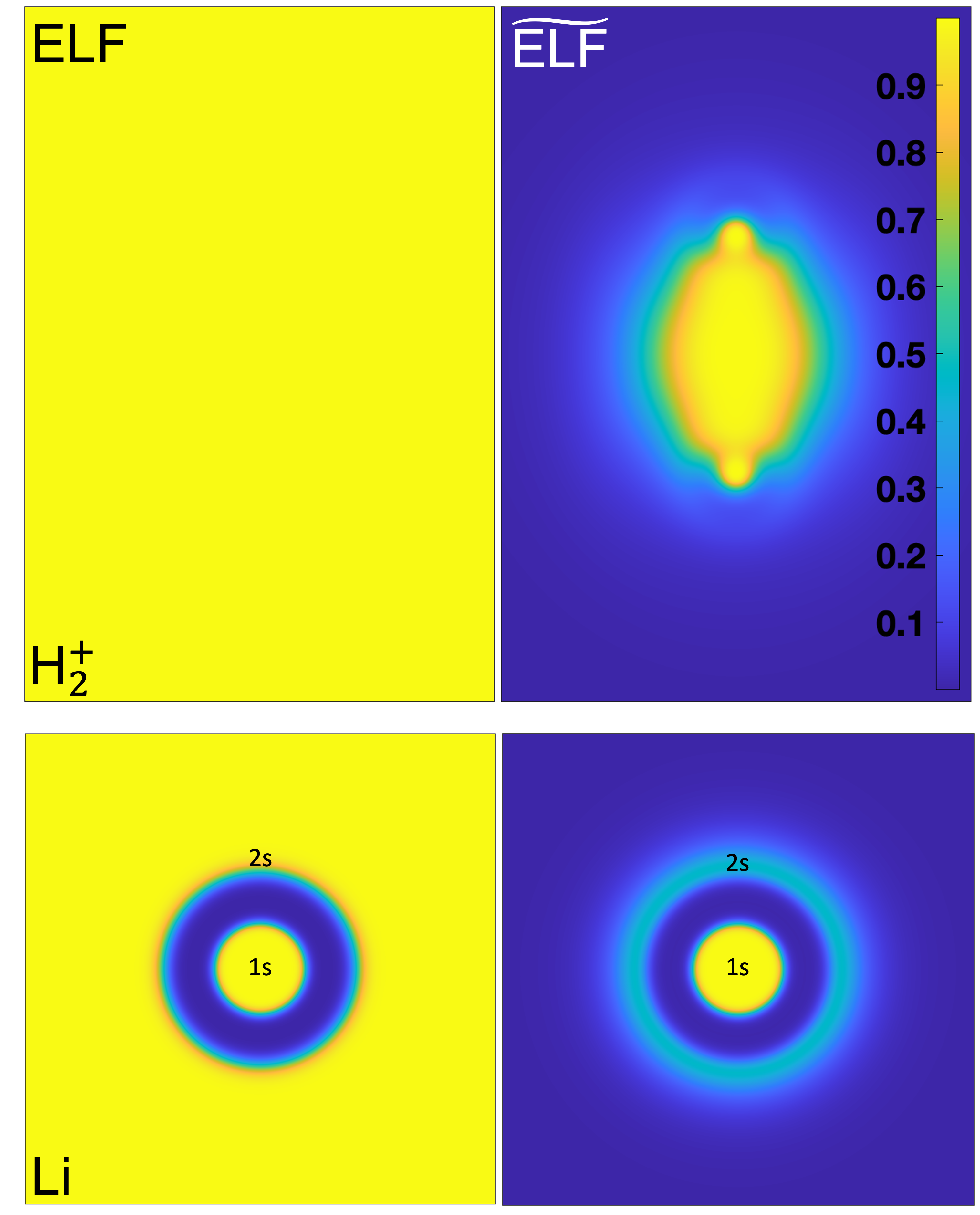}
\caption{(left) ELF and (right) $\widetilde{\mathrm{ELF}}$ for (top) the numerically exact wavefunction of the H$_2^+$ molecule and (bottom) the Li atom. The H atoms are centered on each tip of the lemon-shaped region in the right panel, at a bond distance of 0.74 \AA. The plot of the Li atom is in a 6$\times$6 \AA$^2$ square.}
\label{fig:lemon}
\end{figure}

Next, let us look into  the remarkable novelties that emerge even at the level of simple collinear states. 
Let us consider the simplest of molecules: H$^+_2$.  The (eigen)states of this molecule are spin polarized single-particle ones. 

For this one-electron system, the interpretation of the ELF as a conditional probability as built from the states of two electrons is not applicable, but the ELF may still be interpreted in terms of kinetic energy differences.\cite{Savin92,Savin94} In H$_2^+$ the kinetic energy density of the real state coincides with its single-orbital limit, yielding $D(\br)=\tau(\br)-\nabla n(\br) \cdot \nabla n(\br)/8n(\br)=0$ and, thus,
  $\mathrm{ELF} (\br) = 1$  (see top left panel of Figure \ref{fig:lemon}), everywhere 
--- not showing any structure for this elementary bond.
 This is fixed by the  $\widetilde{\mathrm{ELF}}$  by accessing the extra information on the spin-polarization,
 that are available in $\widetilde{D}(\br)$.
 Figure \ref{fig:lemon} (top right) reports the plot of $\widetilde{\mathrm{ELF}}$, which shows that the electron is shared between the two nuclei.
 
The ELF probes the electron {\em localization} and therefore can capture inhomogeneous electronic structures beyond bonds --- for instance, atomic shells.
In the case of the Li atom (bottom panels of Figure \ref{fig:lemon}), both the ELF and $\widetilde{\mathrm{ELF}}$ display a shell structure (labels for the closed 1$s$ and open 2$s$ shells are provided to guide the eye). 
However the  ELF (left bottom of Figure \ref{fig:lemon}),  predicts large values of the electron localization  also in regions far away from the atomic center where the electrons is unlikely to be found.  Again, the $\widetilde{\mathrm{ELF}}$  (right bottom of Figure \ref{fig:lemon}) fixes the issues by accessing extra information on the spin polarization.\cite{note_asymptotic}

Finally, let us consider the H$_2$ molecule (not plotted). As a consequence of the Pauli principle,  two electrons are in the same
singlet spin configuration at each point in space. Moreover $\vm(\br)=0$, $\vec{\tau}(\br)=0$, and $\vec{\bJ}(\br)=0$ everywhere. 
Hence, $\widetilde{\mathrm{ELF}} (\br)\equiv \mathrm{ELF} (\br)= 1$. A resolution of the localization of the electron pair in H$_2$, would require to account for correlations beyond single Slater determinants\cite{Kohout04,Kohout05,Silvi2010,pittalis2018bonds} --- a task that is beyond the scope of a ``simple'' ELF.

{\em Conclusions}.~ In solving the longstanding issue of extending the electron localization function (ELF) to general non-collinear states, we have discovered that a fundamental step had been so far missed. Namely, the key step is to enforce the  {\em conjoint} U(1)$\times$SU(2) gauge-invariance. This is necessary to ensure that
the  ELF will depend only on intrinsic physical features rather than on {\em any} particular gauge in which the Pauli equation is solved.

The U(1)$\times$SU(2) gauge-invariance adds to an otherwise gauge dependent   ELF --- not only the already known quadratic terms in the particle and spin currents but also ---  a  quadratic term in the gradient of the spin-magnetization {\em and} a term involving the contraction of the magnetization with the (non-collinear) spin-kinetic energy-density. 
These are quantities which, in general, have sizable magnitude that cannot be ignored. 

As illustrations, we have shown the cases of the Cr$_3$ molecule and EuO crystal.
Furthermore, we have shown that, the U(1)$\times$SU(2) gauge-invariant ELF  also improves the description of the localization of one-electron-like collinearly spin-polarized states.

Besides extending the analyses of the electron localization to a wider class of quantum states, in view of the role played by the  ELF (and its components) in the construction of advanced density functional approximations, the extended ELF presented here will help avoid fundamental failures of state-of-the-art density functionals beyond their present domain of applicability.
\\

\begin{acknowledgments}
This research has received funding from the Project CH4.0 under the MUR program ``Dipartimenti di Eccellenza 2023-2027'' (CUP: D13C22003520001). GV was supported by the Ministry of Education, Singapore, under its Research Centre of Excellence award to the Institute for Functional Intelligent Materials (I-FIM, project No. EDUNC-33-18-279-V12). We are grateful to Drs. Jefferson Maul and Chiara Ribaldone for stimulating discussions.
\end{acknowledgments}

%\bibliography{GKS-SCDFT,ELFLIB,JSCAN}

\begin{thebibliography}{66}
\bibitem[{not({\natexlab{a}})}]{note_einstein}
\bibinfo{note}{Here and elsewhere we use Einstein summation over repeated
  indices, unless explicitly specified.}
\bibitem[{cr3()}]{cr3_note}
\bibinfo{note}{The spin-magnetization and spin-kinetic energy-density terms
  have a contribution to the $\widetilde{\mathrm{ELF}}$ on the order of
  $10^{-1}$ a.u., while the spin-currents contribute on the order of $10^{-3}$
  a.u., and the particle currents $10^{-5}$ a.u.}
\end{thebibliography}

\end{document}